\documentclass[twocolumn,aps,prl,showpacs,groupedaddress]{revtex4}
\usepackage{epsfig,amssymb,amsmath}
\def\comment#1{}
\def\togli#1{}
\def\labell#1{\label{#1}}

\begin{document} 

\title{Quantum random access memory}
\author{Vittorio Giovannetti$^1$, Seth Lloyd$^2$, Lorenzo Maccone$^3$}
\affiliation{$^1$NEST-CNR-INFM \& Scuola Normale
  Superiore, Piazza dei Cavalieri 7, I-56126, Pisa, Italy.\\$^{2}$MIT,
  RLE and Dept. of Mech. Engin. MIT 3-160, 77 Mass.~Av., Cambridge, MA
  02139, USA.\\ $^3$QUIT, Dip.  Fisica ``A.  Volta'', Univ. Pavia, via
  Bassi 6, I-27100 Pavia, Italy.}

\begin{abstract}
  A random access memory (RAM) uses $n$ bits to randomly address
  $N=2^n$ distinct memory cells. A quantum random access memory (qRAM)
  uses $n$ qubits to address any quantum superposition of $N$ memory
  cells. We present an architecture that exponentially reduces the
  requirements for a memory call: $O(\log N)$ switches need be thrown
  instead of the $N$ used in conventional (classical or quantum) RAM
  designs. This yields a more robust qRAM algorithm, as it in general
  requires entanglement among exponentially less gates, and leads to
  an exponential decrease in the power needed for addressing. A
  quantum optical implementation is presented.
\end{abstract}
% Quantum computation,  Entanglement and quantum nonlocality, Quantum information
\maketitle 

A fundamental ability of any computing device
is the capacity to store information in an array of memory
cells~\cite{turing}. The most flexible architecture for memory arrays
is random access memory, or RAM, in which any memory cell can be
addressed at will~\cite{ram}.  A RAM is composed of a memory array, an
input register (``address register''), and an output register. Each
cell of the array is associated with a unique numerical address. When
the address register is initialized with the address of a memory cell,
the content of the cell is returned at the output register
(``decoding''). Just as RAM forms an essential component of classical
computers, quantum random access memory, qRAM, will make up an
essential component of quantum computers, should large quantum
computers eventually be built.  It has the same three basic components
as the RAM, but the address and output registers are composed of
qubits (quantum bits) instead of bits. [The memory array can be either
quantum or classical, depending on the qRAM's usage]. The qRAM can
then perform memory accesses in coherent quantum
superposition~\cite{chuang}: if the quantum computer needs to access a
superposition of memory cells, the address register $a$ must contain a
superposition of addresses $\sum_j\psi_j|j\rangle_a$, and the qRAM
will return a superposition of data in a data register $d$, correlated
with the address register:
\begin{eqnarray}
\sum_j\psi_j|j\rangle_a\quad\stackrel{\mbox{\tiny qRAM}}{\longrightarrow}
\quad\sum_j\psi_j|j\rangle_a|D_j\rangle_d
\labell{qram}\;,
\end{eqnarray}
where $D_j$ is the content of the $j$th memory cell. The possibility
of efficiently implementing these devices would yield an exponential
speedup for pattern recognition algorithms
\cite{pattern1,pattern2,pattern3}, period finding, discrete logarithm
and quantum Fourier transform algorithms over classical data.
Moreover, qRAMs are required for the implementation of various
algorithms, such as quantum searching on a classical
database~\cite{chuang}, collision finding~\cite{algorithm1},
element-distinctness in the classical~\cite{algorithm2} and
quantum~\cite{algorithm3} settings, and the quantum algorithm for the
evaluation of general NAND trees~\cite{algorithm4}. Finally, qRAMs
permit the introduction of new quantum computation primitives, such as
quantum cryptographic database searches~\cite{PQQ} or the coherent
routing of signals through a quantum network of quantum
computers~\cite{qrouter}.

Both classical and quantum RAMs are computationally expensive: If the
memory array is disposed in a $d$-dimensional lattice, conventional
architectures involve throwing $O(N^{1/d})$ switches (i.e.~two-body
interactions) to access one out of the $N=2^n$ memory slots, where $n$
is the number of bits in the address register~\cite{ram}.  This
exponential use of resources translates into a relatively slow speed
and high energy usage for classical RAMs during decoding, and to a
high decoherence rate for qRAMs.  For this reason, up to now little
attention has been devoted to developing a qRAM.  In this paper we
introduce a new RAM architecture, dubbed ``bucket-brigade,'' that
reduces the number of switches that must be thrown during a RAM call,
quantum or classical, from $O(N^{1/d})$ to $O(\log N)$\comment{Maybe
  $O(n/d)$?}. If we neglect the travel time of the signals along the
wires connecting the device's components, this translates into an
exponential reduction in the running-time computational-complexity at
the information theoretical level, when compared to conventional
setups. As will be shown, for qRAMs it entails an exponential
reduction in the number of gates that need to be entangled for each
memory call, simplifying the qRAM circuit with respect to the
conventional architectures~\cite{chuang}, and reducing the need for
expensive error correction routines. In addition, the reduction in the
number of switchings translates into a reduction of the energy
employed in the routing, which may yield more efficient RAMs that use
less power during decoding than current architectures.

We start by describing the conventional RAM architecture, showing why
its direct translation to the quantum realm is inefficient and
noise-prone. We then introduce our bucket-brigade architecture and
give an account of the required resources in the classical and quantum
setting. We conclude introducing an illustrative example.

\begin{figure}[h!]
\begin{center}
\epsfxsize=1.\hsize\leavevmode\epsffile{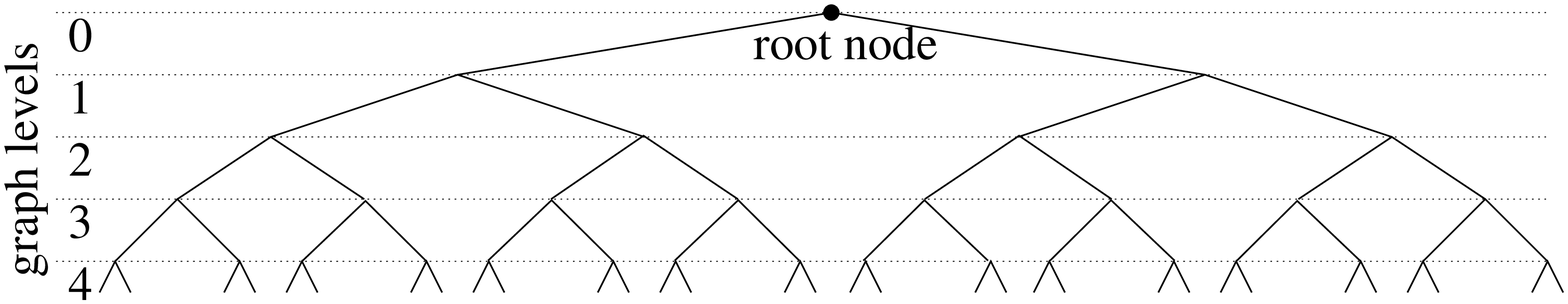}
\end{center}
\vspace{-.5cm}
\caption{Bifurcation graph of the RAM addressing.  }
\label{f:fig}\end{figure}

\paragraph*{Quantum RAM}
Even though more elaborate architectures exist~\cite{ram} (such as
ones using $d$-dimensional memory arrays), the basic RAM addressing
scheme is simple: Suppose that the $N$ memory cells are placed at the
end of a bifurcation graph, composed by the $n$ levels shown in
Fig.~\ref{f:fig}.  The value of the $j$th bit in the address register
can be interpreted as the route to follow when one has reached a node
in the $j$th level of the graph: if the value is 0, the left path must
be followed; if it is 1, the right path must be followed (e.g.~an
address register 010..  is interpreted as `left at the 0th level,
right at the first level, left at the second', etc.).  Each of the $N$
possible values of the address register thus indicates a unique route
that crosses the whole graph and reaches one of the memory
cells~\cite{nota}.  An electronic implementation requires placing one
transistor in each of the two paths following each node in the graph.
Each address bit controls all the transistors in one of the graph
levels: it activates all the transistors in the left paths if it has
value 0, or all the transistors in the right paths if it has value
1~\cite{ram}.  Thus, an exponential number of transistors must be
activated at each memory call to route the signals through the graph
(this entails an energy cost exponentially larger than the cost of a
single transistor activation.)

Direct translations of the above scheme into the quantum
realm~\cite{chuang} are quite impractical. The $n$ qubits of the
address register coherently control $n$ quantum control lines, each of
which acts coherently on an entire level of the bifurcation graph.  At
each branch of the bifurcation graph, a 0 in the address register for
that level shunts signals along the left paths, and a 1 shunts signals
along the right paths.  Each binary address is correlated with a set
of switches that pick out the unique path through the graph associated
with that address.  A coherent superposition of addresses is
coherently correlated, i.e.~entangled, with a set of switches that
pick out a superposition of paths through the graph.  To complete the
quantum memory call, a quantum `bus' is injected at the root node and
follows the superposition of paths through the graph. Then the
internal state of the bus is changed according to the quantum
information in the memory slot at the end of the paths (e.g.~through a
controlled-NOT transformation that correlates the bus and the
memory)~\cite{nota1}.  Finally, in order to decorrelate the bus
position from the address register, the bus returns to the root node
by the same path.  Like a quantum particle, the bus must be capable of
traveling down a coherent superposition of paths.  Although not
impossible, such a qRAM scheme is highly demanding in practice for any
reasonably sized memory.  In fact, to query a superposition of memory
cells, the address qubits are in general entangled with $O(N)$
switches or quantum gates (or, equivalently, they must control
two-body interactions over exponentially large regions of space), i.e.
a state of the form $\sum_j\psi_j\ |j_0j_1\cdots
j_{n-1}\rangle_a\otimes|j_0\rangle_{s_0}|j_1\rangle_{s_1}^{\otimes 2}
\cdots|j_{n-1}\rangle_{s_{n-1}}^{\otimes 2^{n-1}}$, where $j_k$ is the
$k$th bit of the address register, and $s_k$ is the state of
the $2^{k}$ switches controlled by it. Such a gigantic superposition
is highly susceptible to decoherence and requires costly quantum error
correction whenever the error rate is bigger than $2^{-n}$.  In fact,
if a single gate out of the $N=2^n$ gates in the array is decohered,
then the fidelity of the state in average is reduced by a factor of
two, and if at least one gate in each of the $k$ lines is decohered,
the fidelity in average is reduced by $2^{-k}$.  \comment{All one
  needs to do is to do a homogeneous average over all input states and
  then decohere one of the qubits: evaluate the fidelity and take the
  average: $\langle\psi|\rho|\psi\rangle= \sum_{J_i,\vec
    J}\sum_{J_i,\vec J'}|\psi_{J_i,\vec J}|^2|\psi_{J_i,\vec
    J'}|^2=1/2^n1/2^{n-1}$, where $J_i$ is the qubit that got
  dephased.}

The ``bucket-brigade'' is based on sending both the address register
and the signal through the bifurcation graph.  Like buckets of water
passed along a line of improvised fire-fighters, they carve a route
that crosses the whole graph along which the information can be
extracted. With respect to the conventional architecture detailed
above, the $O(N)$ active logic gates are replaced by memory
elements, most of which are in a passive $wait$ state during each
memory call. As a result, there is an exponential reduction of active
gates and of two-body interactions, from $O(N)$ to $O(\log^2N)$.  This
means the bucket-brigade RAM could also be useful in classical
computation to reduce the energy needed for the addressing. (Hybrid
schemes that combine the two above architectures might be more
generally useful).

\begin{figure}[h!]
\begin{center}
\epsfxsize=.9\hsize\leavevmode\epsffile{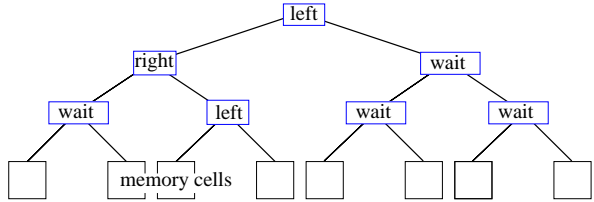}
\end{center}
\vspace{-.5cm} \caption{Bifurcation graph of the bucket-brigade
  architecture. Here the third memory cell is addressed (address
  register 010).}
\label{f:fbb}\end{figure}

The basic idea follows. At each node of the graph of Fig.~\ref{f:fig}
there is a trit, a three-level memory element.  The trit's three
levels are labeled $wait$, $left$, and $right$. A trit in the level
$wait$ will change its value according to the value of any incoming
bit: if the incoming bit is 0, it takes the value $left$, while if the
incoming bit is 1, it takes the value $right$. A trit in the level
$left$ or $right$ will deviate any incoming signal along the graph
according to its value. The protocol starts by initializing all the
trits in the state $wait$. Then the first bit of the address register
is sent through the graph. It will induce a change in the root node,
which will be transferred to $left$ or $right$ depending on the bit's
value. Now the second bit of the address register is sent through the
graph. Depending on the value of the first node, it will be deviated
left or right and will meet one of the two nodes on the second level
of the graph (both of which are in a $wait$ state).  This node will be
transformed according to the bit's value, and so on.  After all the
$\log N$ bits of the address register have passed through the graph, a
single route of $n=\log N$ $left$-$right$ trit states has been carved
through the graph (see Fig.~\ref{f:fbb}).  All other trits remain in
the $wait$ state. Now a {\em bus} signal can easily follow this route
(by heeding the indications of the trits it encounters) and find its
way to the element in the memory array that the address register was
pointing to. Information is then extracted through this route by
sending back the bus signal, which must again heed the directions of
the trits it encounters while traveling to the graph's root node. In
addition, every time the bus signal on its way back encounters a trit,
the trit is reset to the $wait$ state.  Thus, the memory element is
addressed by the bus signal, which is then sent back to the root node,
and the graph is reset to its initial $wait$ state. Only $\log N$
trits have been involved in the memory call.

In the quantum realm the trits must be replaced by qutrits,
i.e.~three-level quantum systems, described by the vectors
$|wait\rangle$, $|left\rangle$, and $|right\rangle$.  Now, when the
qubits of the address register are sent through the graph, at each
node they encounter a unitary encoding transformation $U$.  If the
qutrit is initially in the $|wait\rangle$ state, the unitary swaps the
state of the qubit in the two $|left\rangle$-$|right\rangle$ levels of
the qutrit (i.e.  $U|0\rangle|wait\rangle=|f\rangle|left\rangle$ and
$U|1\rangle|wait\rangle=|f\rangle|right\rangle$, where $|f\rangle$ is
a fiduciary state of the qubit). If the qutrit is not in the
$|wait\rangle$ state, then it simply routes the incoming qubit
according to its state.  It is clear that an address register in a
quantum superposition will carve a superposition of routes through the
graph, so that any incoming qubit will exit the graph in the
corresponding superposition of locations. Once all the register qubits
are sent through the graph, a bus qubit is injected and it reaches the
end of the graph along the requested superposition of paths. It then
interacts with the memory cells at such locations changing its state
according to their information content.  Now the bus qubit is sent
back through the graph, exiting at the graph's root node.  Finally,
starting from the last level of the graph, the qutrits are subject to
the inverse of the unitary encoding transformation: a qutrit initially
in the states $|left\rangle$ or $|right\rangle$ is evolved back to the
state $|wait\rangle$, while sending a qubit (containing the state of
the $|left\rangle$-$|right\rangle$ levels) back through the graph,
i.e.  the transformation
$U^\dag|f\rangle|left\rangle=|0\rangle|wait\rangle$ or
$U^\dag|f\rangle|right\rangle=|1\rangle|wait\rangle$. In order to
activate this transformation at the right moment, various schemes are
possible. The simplest one entails activating a classical control over
all the qutrits in each level of the tree, sequentially from the last
level up to the root node. Alternatively, one can send $n$ control
qubits along the superposed path, each of which controls the unitary
$U^\dag$ at one of the tree levels. A further scheme entails
introducing counters in each node, which activate the $U^\dag$ unitary
after a level-dependent number of signals have transited. At the end,
all qubits of the address register have been ejected from the graph,
which is restored to its initial state of all qutrits in the
$|wait\rangle$ state, yielding the transformation of Eq.~(\ref{qram}).

Similarly to what happens in quantum computation with atomic
ensembles~\cite{deco}, the noise resilience of the bucket-brigade
stems from the fact that in each branch of the superposition only
$\log N$ qutrits are not in the passive $|wait\rangle$ state. In fact,
for a query with a superposition of $r$ memory cells, it is necessary
to entangle only $O(r\log N)$ qutrits, as the state of the device is
of the type
  $\sum_j\psi_j|j_0\rangle_{{t}_0}|j_1\rangle_{{t}_1({j_0})}
  \cdots|j_{n-1}\rangle_{{t}_{n-1}({j_{n-2}})}\otimes_{\ell_j}
  |wait\rangle_{{t}_{\ell_j}}$,
where ${t}_k$ represents the state of the one qutrit at the $k$th
level which is aimed to by the non-$wait$ qutrit at the $k-1$ level,
and where $\ell_j$ spans the other qutrits. Even if all of the qutrits
are involved in the superposition, the state is still highly resilient
to noise: if a fraction $\epsilon$ of the gates are decohered (with
$\epsilon \log N < 1$) then in average the fidelity of the resulting
state is $O(1-\epsilon \log N)$ (compare this to the 1/2 fidelity
reduction in the conventional qRAM above). The noise resilience is, of
course, greater in those algorithms where $r$ is small, such as the
QPQ~\cite{PQQ} or the quantum routing~\cite{qrouter}. Moreover, note
that the exponentially larger number of $|wait\rangle$ states could
give significant overall errors even if their individual error rates
are much lower than those used in the left and right states.

\begin{figure}[h!]
\begin{center}
\epsfxsize=.65\hsize\leavevmode\epsffile{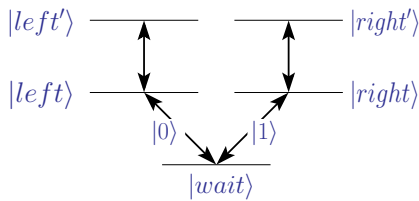}
\end{center}
\vspace{-.5cm} \caption{Basic level structure of the atoms in a
  possible bucket-brigade implementation. Some extra energy levels
  needed to implement Raman transitions are not shown.}
\label{f:struct}\end{figure}

\paragraph*{Bucket-brigade implementation}
Like cluster state quantum computation~\cite{cluster}, the
bucket-brigade only assumes the possibility of operating coherently on
a small number $O(\log N)$ out of large number $O(N)$ of
first-neighbor connected quantum memory elements, and it does not
require macroscopic superposition states composed of an exponentially
large number of quantum gates.  Candidate systems for bucket-brigade
qRAMs include optical lattices~\cite{optical,optical1}, Josephson
arrays~\cite{romito}, arrays of coherently coupled quantum dots, or
strongly correlated cavity arrays~\cite{hart}.  In order to be more
specific on the nature of the necessary resources, we present a
proof-of-principle implementation of the quantum bucket-brigade. (It
should be only considered as an illustrative example, and not as an
experimental proposal. More detailed versions of bucket-brigade
implementations will be presented in future work.)  The qutrits at the
nodes of the graph of Fig.~\ref{f:fig} are composed of trapped atoms
or ions with the level structure depicted in Fig.~\ref{f:struct}: a
ground state $|wait\rangle$ and two excited states $|left\rangle$ and
$|right\rangle$. The register and bus qubits are composed by photons,
whose encoding is in the polarization. 
It is now possible to use a photon in the polarization state
$|0\rangle$ to muster a $|wait\rangle\to|left\rangle$ atomic
transition, and a photon in the polarization state $|1\rangle$ to
muster a $|wait\rangle\to|right\rangle$ transition. Furthermore by
employing Raman techniques, one use strong classical pulses that
couple $|wait\rangle$, $|left\rangle$ and $|right\rangle$ with
extra energy levels (not shown in the picture) to externally control
the timing of such transitions. Note that, being classical, such
pulses do not need to act locally on a single atom but they can
interact with all the nodes of each level.
Thus, a photon impinging on an atom in the $|wait\rangle$ state
transfers its internal state to the $|left\rangle$-$|right\rangle$
atomic levels.  A photon impinging on an atom which is in a
$|left\rangle$ state, will excite a cyclic transition (using the level
$|left'\rangle$) and is re-emitted by the atom. The
$|left\rangle\to|left'\rangle$ transition is insensitive to the
photon's polarization and is coupled to an outgoing spatial mode
departing the trapped atom in the left direction. This means that a
photon in any polarization state that impinges onto an atom in the
$|left\rangle$ state is deviated along the graph towards the left.
Analogously, a photon in any state impinging on an atom in the
$|right\rangle$ state is deviated towards the right. As in the
$|wait\rangle\rightarrow |left,right\rangle$ transition, the timing
of the whole process can be controlled by coupling the involved states
with ancillary levels through strong classical Raman pulses.  After
all the photons of the address register are sent through the graph, a
bus photon (initially in the state $|0\rangle$) is injected.  Thanks
to the above mechanism, it crosses the graph in a coherent
superposition of paths, exiting at the location of the addressed cells
and changing its polarization state according to their memory content.
It is then reflected back through the graph and is again deflected
interacting with the atoms, so that it exits the graph at the root
node. To end the protocol, the Raman process is inverted, step by
step, starting from the last level in the graph, so that the atomic
levels $|left\rangle$ and $|right\rangle$ are driven to the
$|wait\rangle$ level, through the emission of a $|0\rangle$ or
$|1\rangle$ photon respectively. Thus the address register photons are
emitted one-by-one and coherently driven back through the graph to the
root node.

\paragraph*{Conclusions} We have described a RAM architecture where
active gates are replaced by three-level memory elements.
It could give rise to a significant simplification in the qRAM
implementation, to exponentially reduced decoherence rate and energy
saving. However, in current RAMs, the primary sources of dissipation
are leakage current in the memory cells (for SRAMs) and refreshing
memory cells (for DRAMs).  Energy costs in the memory access procedure
are not currently important enough to warrant accepting the additional
delays and memory elements of the bucket-brigade.  For future,
non-CMOS RAMS, however, decoding energy costs may become important, so
that the exponential savings of the bucket-brigade architecture may
prove significant.

%In addition to aiding in the construction of large quantum computers,
%our setup would constitute an enabling technology for a wide range of
%quantum information processes (e.g.~pattern
%recognition~\cite{pattern1,pattern2,pattern3}), which use the qRAM for
%coherently representing data in order to provide exponential speedups
%over classical algorithms.

\acknowledgments{We acknowledge useful feedback from A. Childs and
  support from Jeffrey Epstein. }


\begin{references}
\bibitem{nota1}We focus on the case in which we ``read'' from the
  memory, the ``write'' operation being completely analogous.
\end{references}
\end{document}